\documentclass[aps,pra,twocolumn,showpacs,floatfix]{revtex4-1}

\usepackage{amssymb,amsmath}
\usepackage{graphicx}
\usepackage{eurosym}
\usepackage{color}
\usepackage[normalem]{ulem}
\usepackage[makeroom]{cancel}
\usepackage{bm}
\usepackage{ulem}

\newcommand{\bpsi}{{\bm \psi}}
\newcommand{\bphi}{{\bm \phi}}
\newcommand{\ks}{{k_\star}}
\newcommand{\ba}{{\bm a}}

\newcommand{\bb}{{\bm b}}

\newcommand{\cM}{\mathcal{M}}

\newcommand{\cC}{\mathcal{C}}

\newcommand{\hU}{\hat{U}}

\newcommand{\tcM}{\tilde{\mathcal{M}}}

\newcommand{\p}{{\cal P}}
\newcommand{\PT}{{\cal PT}}
\newcommand{\T}{{\cal T}}

\newcommand{\K}{{\cal K}}

\newcommand{\cK}{{ { \cal K}}}
\newcommand{\cN}{{ { \cal N}}}

\newcommand{\tuu}{t_{\uparrow\uparrow}}
\newcommand{\tud}{t_{\uparrow\downarrow}}
\newcommand{\tdu}{t_{\downarrow\uparrow}}
\newcommand{\tdd}{t_{\downarrow\downarrow}}
\newcommand{\ruu}{r_{\uparrow\uparrow}}
\newcommand{\rud}{r_{\uparrow\downarrow}}
\newcommand{\rdu}{r_{\downarrow\uparrow}}
\newcommand{\rdd}{r_{\downarrow\downarrow}}

\newcommand{\up}{|\!\! \uparrow \rangle}
\newcommand{\dn}{|\!\! \downarrow \rangle}

\begin{document} 

\title{Spectral singularities of odd-$\PT-$symmetric potentials}

\author{Vladimir V. Konotop$^1$ and Dmitry A. Zezyulin$^{2}$}
 \affiliation{$^1$Departamento de F\'isica and Centro de F\'isica Te\'orica e Computacional, Faculdade de Ci\^encias, Universidade de Lisboa, Campo Grande 2, Edif\'icio C8, Lisboa 1749-016, Portugal
 	\\
 	$^2$ITMO University, St.~Petersburg 197101, Russia 
}

\begin{abstract} 
 We describe one-dimensional stationary scattering of  a two-component wave field by a non-Hermitian matrix potential which features odd-$\PT$ symmetry, i.e., symmetry with $(\PT)^2=-1$. The scattering is characterized  by a $4\times 4$ transfer matrix. The main attention is focused on spectral singularities which are classified into two types. \textit{Weak} spectral singularities are characterized by zero determinant  of a diagonal $2\times 2$ block of the transfer matrix. This situation corresponds to the lasing or coherent perfect absorption of pairs of oppositely polarized modes.
 {\em Strong} spectral singularities  are characterized by  zero diagonal block of the transfer matrix. We show that in   odd-$\PT$-symmetric systems  any spectral singularity is self-dual, i.e., lasing and coherent perfect absorption  occur simultaneously. Detailed analysis is performed for a case example of a  $\PT$-symmetric coupler consisting of two  waveguides, one with localized gain and another with localized absorption, which are coupled by a localized antisymmetric medium. For this coupler, we discuss  weak self-dual spectral singularities and their splitting into complex-conjugate eigenvalues which represent bound states characterized by  propagation constants with real parts lying in  the continuum. A rather counterintuitive restoration of the unbroken odd-$\PT$-symmetric phase subject to the increase of the gain-and-loss strength is also revealed. The comparison between odd-- and even--$\PT-$symmetric couplers, the latter characterized by $(\PT)^2=1$, is also presented.
\end{abstract}

\maketitle

\section{Introduction}

Spectral singularities (SSs) in  continuous spectra of  
non-Hermitian operators are defined as real poles of the resolvent. In the mathematical literature they have been  
known  for several decades~\cite{SS}.  
More recently, SSs have attracted considerable attention in the physical literature, where they represent wavenumbers at which the reflection and transmission coefficients become infinite~\cite{Mostafazadeh2009,ScarfII}. If a SS is a real positive pole of the resolvent, it allows for a coherent lasing solution~\cite{Longhi,rectang}. If a SS corresponds a negative wavenumber corresponds to the coherent perfect absorption (CPA). A CPA solution can also be viewed as time-reversed lasing  \cite{Stone,reviewCPA}, and therefore such a SS is often referred to as a time-reversed SS.  When a system obeys parity  ($\p$) - time ($\T$) symmetry~\cite{BenderBoet}, a SS and time-reversed SS appear at the same wavelength, i.e., at the same value of the modulus of the wavenumber \cite{Longhi,Stone_selfdual}. This means that in a $\PT$-symmetric system any spectral singularity is \textit{self-dual}, and
the system  can operate   as a laser and as a coherent perfect absorber simultaneously at the given wavelength. 
Apart from $\PT$-symmetric potentials, there exist other special situations where self-dual spectral singularities also occur \cite{rectang,KZ2017}.  

A non-Hermitian Hamiltonian with a SS is non-diagonalizable~\cite{Mostafazadeh2009JPA}. This  resembles a similar property of Hamiltonians featuring exceptional points \cite{Kato} which correspond to the situation
where two  eigenvalues from the discrete spectrum and the corresponding eigenvectors coalesce subject to the change of some control  parameter of the system. The analogy between SSs and exceptional points can be extended further, since both these phenomena may correspond to the transition from purely real to complex spectra of the Hamiltonian (also known as the transition from unbroken to broken $\PT$-symmetric phase). Indeed, it is well-known \cite{PTbreaking}  that, subject to the change  of some control parameter,  a $\PT$-symmetric system can be driven to  an  exceptional point, where two  isolated eigenvalues coalesce and form a double eigenvalue  associated with a Jordan block (which means that there is only one linearly independent eigenfunction associated with the double eigenvalue). Past the exceptional point, the double  eigenvalue splits into a pair of simple complex conjugate eigenvalues.
In a similar way, a non-Hermitian system  can  undergo the transition  from a purely real to complex spectrum
if changing  some of its parameters triggers  a  self-dual spectral singularity in the continuous spectrum \cite{KZ2017}.  Past the self-dual spectral singularity, a pair of complex-conjugate eigenvalues emerges from an interior point of the continuous spectrum. This represents an alternative scenario of the $\PT$-phase breaking~\cite{Yang,KZ2017}. 

An antilinear time-reversal operator $\T$ can be implemented in physically different ways~\cite{Messiah}. The first one corresponds to the conventional ``bosonic'' time reversal   characterized by the property $\T^2=1$. This is the most used $\T$ operator in numerous applications of $\PT$ symmetry in quantum mechanics~\cite{BenderBoet} and in other fields \cite{KZYreview}. The description of the state of the art given above refers precisely to this case, which  will  be also termed as {\em even}-$\PT$ symmetry in what follows. In the meantime, an alternative ``fermionic'' time reversal, characterized by  the property  $\T^2=-1$, has also been introduced in the theory of non-Hermitian systems~\cite{SmithMathur,BendKlev}, but so far  it has received much less attention. Recently, we have shown~\cite{ZK2018} that {\em odd}-$\PT$ symmetry (we use this term to distinguish the $\PT$ symmetry with $\T^2=-1$ from the conventional even  $\PT$ symmetry   with $\T^2=1$) naturally appears in modeling  wave propagation in combined 
$\PT-$symmetric and anti-$\PT$-symmetric~\cite{Ge} media. The odd-$\PT$-symmetric  model    introduced in~\cite{ZK2018} was an example of a discrete optical system, allowing one to explore the effects related to the intrinsic degeneracy of the discrete spectrum, including $\PT-$symmetry breaking through exceptional points. 

The goal of the present work is twofold. First, we introduce an odd-$\PT$-symmetric waveguiding system which takes into  account   diffraction (or dispersion) of waves.  Second, we describe the scattering by a localized odd-$\PT$-symmetric potential, focusing on the emergence of spectral singularities. 

The paper is organized as follows. The model is introduced in Sec.~\ref{sec:model}. In Sec.~\ref{sec:formalism} we address some general characteristics of scattering of two-component fields and discuss two types of SSs. Analysis of the properties of the transfer matrix and SSs in a system obeying odd-$\PT$ symmetry is presented in Sec.~\ref{sec:SS}. In Sec.~\ref{sec:example} we perform detailed analysis of the stationary scattering by a localized potentials in an odd-$\PT$-symmetric coupler and compare the results with those obtained for scattering by conventional (even-)$\PT$-symmetric potentials. The results are summarized and discussed in the last Sec.~\ref{sec:conclusion}.

\section{The model}
\label{sec:model}

We consider the one-dimensional stationary scattering problem for a two component field $\bpsi=(\psi_1,\psi_2)^T$ (hereafter the upper index ``$T$'' means ``transpose'')
\begin{eqnarray}
\label{eq:scat_prob}
H\bpsi=k^2\bpsi,
\end{eqnarray}
 where $k$ is the spectral parameter and the Hamiltonian is given by
\begin{equation}
\label{Hamiltonian}
H=-\frac{d^2}{dx^2}\sigma_0+\hat{U}, 
\quad 
\hat{U}=\left(\begin{array}{cc}
U_0(x)& V_2(x) \\
V_2(x)& U_1(x)
\end{array}
\right).
\end{equation}
Here $\sigma_0$ is the $2\times 2$ identity matrix, and all entries of the complex-valued matrix potential $\hat{U}$ vanish at the infinity:
\begin{eqnarray}
\label{eq:limits}
\lim_{x\to\pm\infty} U_{0,1}(x)=0, \qquad \lim_{x\to\pm\infty} V_{2}(x)=0.
\end{eqnarray}
 
In order to impose specific conditions on the  matrix potential $\hat{U}(x)$, we first recall the definitions of the space inversion, $\p$, and odd-time-reversal, $\T$, operators~\cite{Messiah}: 
\begin{eqnarray}
 \p:\, x\to-x,
 \quad  \T=\sigma_2\K.
\end{eqnarray}
 Hereafter $\sigma_{1,2,3}$ are the Pauli matrices, and $\K$ is the conventional  elementwise complex conjugation:  $\cK\bpsi=\bpsi^*$ (hereafter an asterisk is used  for complex conjugation, too). Obviously, $\p^2=1$,  $\T^2=-1$, $[\p, \T]=0$, and hence $(\PT)^2=-1$ [in contrast to   more often used  even parity-time reversals for  which   $(\PT)^2=1$].  We require  Hamiltonian $\hat{U}$ to be odd-$\PT$ symmetric, i.e., to satisfy 
\begin{equation}
\label{PT}
[H,\PT]=0,
\end{equation}
which is equivalent to the conditions 
\begin{equation}
\label{eq:cond}
U_0(x) = U_1^*(-x), \quad V_2(x) = -V_2^*(-x).
\end{equation} 
 
The latter condition requires the coupling $V_2(x)$ to be anti-$\PT$-symmetric, with the even time-reversal operator being simply the complex conjugation $\cK$~\cite{Ge}, which  corroborates with the coupled-waveguide model introduced in \cite{ZK2018}. At the same time, as shown below in Sec.~\ref{sec:example}, even a   simpler case of 
real-valued $V_2(x)$   leads to nontrivial results.

\section{General formalism}
\label{sec:formalism}

\subsection{Jost solutions, transfer matrix, scattering coefficients}
 
First, we introduce some general definitions and scattering characteristics for the spinor system  (\ref{eq:scat_prob})-(\ref{Hamiltonian}), which will be used below.  No specific symmetry of the matrix-potential $\hU(x)$ is assumed so far, i.e., condition  (\ref{PT}) is not imposed  in this section. It is  yet assumed that all entries of $\hU(x)$  are localized [i.e., Eqs.~(\ref{eq:limits}) hold] and tend to zero fast enough, such that the continuous spectrum of scattering states occupies the real semiaxis  $k^2\geq 0$.  As usually, the Jost solutions of the scattering problem (\ref{eq:scat_prob})--(\ref{Hamiltonian})  are  defined  by their asymptotics at $x\to-\infty$ (denoted by $j=1$) and at $x\to\infty$ (denoted by $j=2$):
\begin{subequations}
	\label{Jost}
\begin{eqnarray}
	\bphi_{1j}(x,k)&\to&  \left(\begin{array}{c}
		1\\0
	\end{array}\right)e^{ikx}=  \up  e^{ikx},
	\\[2mm]
	\bphi_{2j}(x,k)&\to&  \left(\begin{array}{c}
		0\\1
	\end{array}\right)e^{ikx}= \dn  e^{ikx},
	\\[2mm]
	\bphi_{3j}(x,k)&\to&  \left(\begin{array}{c}
		1\\0
	\end{array}\right)e^{-ikx}= \up e^{-ikx},
	\\[2mm]
	\bphi_{4j}(x,k)&\to&  \left(\begin{array}{c}
		0\\1
	\end{array}\right)e^{-ikx}= \dn e^{-ikx}.
\end{eqnarray}
\end{subequations}
The vectorial character of the two-component solutions  makes it convenient to   use  the terminology of the up--polarized, $\up$, and down--polarized, $\dn$, states. 

While we are primarily interested in the behavior of the system at  real wavenumbers $k$, the Jost solutions   can be defined in the complex $k$-plane by the analytic continuation from the real axis.

The $4\times 4$ transfer matrix  $M(k)$ and its inverse $M^{-1}(k)$ 
are  introduced through the relations
\begin{eqnarray}
\label{def_M}
\bphi_{j1}(x,k)=\sum_{m=1}^4 M_{mj}(k)\bphi_{m2}(x,k),
\\
\label{def_tM}
\bphi_{j2}(x,k)=\sum_{m=1}^4 (M^{-1})_{mj}(k)\bphi_{m1}(x,k).
\end{eqnarray}
Using  standard arguments of the ODE theory,  one can  prove that
\begin{equation}
\label{eq:detM1}
\det M(k) = 1.
\end{equation}

From the definition of the Jost functions, it is clear that both transition and reflection of an incident polarized wave can occur with, $\alpha\to\beta$, and without, $\alpha\to\alpha$, inversion of polarization  (here $\alpha$ and $\beta$ stand for the states $\uparrow$ and $\downarrow$). Respectively, we introduce reflection, $r_{\alpha\beta}$, and transmission, $t_{\alpha\beta}$, coefficients, which are  defined through the asymptotics of the solutions $\bphi_\alpha^{L,R}(x)$, where the upper indexes $L$ and $R$ stand for left and right incidence, respectively. To define such solutions it is convenient to consider only positive $k>0$ and identify the left $e^{ikx} |\alpha\rangle $ and right  $e^{- ikx} |\alpha\rangle $  incident waves. Respectively we have:
\begin{eqnarray*}
	\label{asymp1}
	\bphi_{\uparrow}^L &\to& \left\{ 
	\begin{array}{ll}
		e^{ikx} \up + \ruu^L e^{-ikx}\up + \rud^L e^{-ikx} \dn, & x\to-\infty,\\[3mm]
		\tuu^L   e^{ikx} \up + \tud^L e^{ikx} \dn, & x\to+\infty.
	\end{array}
	\right.
	\\
	\label{asymp2}		\bphi_{\downarrow}^L &\to& \left\{ 
	\begin{array}{ll}
		e^{ikx} \dn + \rdd^L e^{-ikx}\dn + \rdu^L e^{-ikx} \up, & x\to-\infty,\\[3mm]
		\tdd^L   e^{ikx} \dn + \tdu^L e^{ikx} \up, & x\to+\infty,
	\end{array}
	\right.
	\\
	\label{asymp3}
	\bphi_{\uparrow}^R &\to& \left\{ 
	\begin{array}{ll}
		e^{-ikx} \up + \ruu^R e^{ikx}\up + \rud^R e^{ikx} \dn, & x\to+\infty,\\[3mm]
		\tuu^R   e^{-ikx} \up + \tud^R e^{-ikx} \dn, & x\to-\infty,
	\end{array}
	\right.
	\\
	\label{asymp4}
	\bphi_{\downarrow}^R &\to& \left\{ 
	\begin{array}{ll}
		e^{-ikx} \dn + \rdd^R e^{ikx}\dn + \rdu^R e^{ikx} \up, & x\to+\infty,\\[3mm]
		\tdd^R   e^{-ikx} \dn + \tdu^R e^{-ikx} \up, & x\to-\infty.
	\end{array}
	\right.
\end{eqnarray*}
Finally, comparing these asymptotics with the definitions (\ref{Jost})--(\ref{def_tM}), we can  express   the reflection and transmission coefficients through the elements of the transfer matrix. These expressions are presented  in Appendix~\ref{app:A}.

 \subsection{Spectral singularities, CPA and lasing}
\label{sec:CPA-lasing}

Now we turn to SSs and their physical implications. 
Since any  scattering state $\bpsi(x)$ from the continuous spectrum can be represented a linear combination  of the Jost solutions, in the    limit $x\to-\infty$ it has the following asymptotics 
\begin{equation}
\label{eq:psiaaaa}
\bpsi(x)\to a_1 e^{ikx} \up + a_2 e^{ikx}\dn + a_3 e^{-ikx} \up + a_4 e^{-ikx}\dn,
\end{equation}
where $a_j$ are some coefficients, and for $k>0$ the terms proportional to $ e^{ikx}$ and $e^{-ikx}$ correspond to the waves propagating towards and outwards the potential, respectively.
Similarly, 
in the limit $x\to\infty$ the same scattering state   behaves as
\begin{equation}
\label{eq:psibbbb}
\bpsi(x) \to b_1 e^{ikx} \up + b_2 e^{ikx}\dn + b_3 e^{-ikx} \up + b_4 e^{-ikx}\dn,
\end{equation}
and  for $k>0$ the terms proportional to $e^{ikx}$ and $ e^{-ikx}$ correspond to the waves propagating outwards and towards the potential, respectively. Next, one observes that  column-vectors
\begin{equation}
\ba=(a_1,a_2,a_3,a_4)^T,\quad  \bb=(b_1,b_2,b_3,b_4)^T,
\end{equation}
 are related by the transfer matrix: 
\begin{equation}
\label{bMa}
\bb=M\ba.
\end{equation}

Since the transfer matrix couples two right-- and left--propagating waves with two polarizations, it is convenient to analyze its block representation  
\begin{eqnarray}
M(k)=\left(\begin{array}{cc}
\cM_{11} (k)& \cM_{12}(k)
\\
\cM_{21}(k) & \cM_{22}(k)
\end{array}\right),
\end{eqnarray}
where $\cM_{ij}$ are $2\times 2$ matrices. Indeed,   the diagonal blocks $\cM_{jj}$ describe total transmission, while the antidiagonal blocks $\cM_{ij}$ ($i\neq j$) describe total reflection.

Formally SSs are determined by properties of the diagonal block $\cM_{22}(k)$ at $k\in\mathbb{R}$. Since however,  $\cM_{22}(-k)=\cM_{11}(k)$ for any real $k$, below we consider SSs only at $k>0$. This is convenient for the analysis of the reflection and transmission coefficients introduced above.  Thus, in this approach, the time reversed SS is determined by the properties of the block $\cM_{11}(k)$ at $k>0$.

It is evident from the formulas for the scattering data shown in Appendix~\ref{app:A}, that their singularities are expected when 
\begin{equation}
\label{Delta}
\Delta (k_\star)=0,  \mbox{ where $\Delta(k):= \det\cM_{22}(k)$,}
\end{equation}
and $k_\star$ is real.

The lasing corresponds to the existence of only outgoing solutions for a given $k=k_\star>0$, i.e., to   solutions $\psi_{LAS}(x)$ whose asymptotic behavior in (\ref{eq:psiaaaa}) and (\ref{eq:psibbbb}) is described by the following column-vectors
%
\begin{eqnarray}
\label{las_solut}
\ba_{LAS}=(0,0,a_3,a_4)^T, \quad  \bb_{LAS}=(b_1,b_2,0,0)^T.
\end{eqnarray}
For such solutions Eq.~(\ref{bMa})  reduces to
\begin{equation}
\label{laser}
\cM_{22}(k_\star)\left(\begin{array}{c}
a_3 \\ a_4
\end{array}\right)=0_{2,1},
\quad   
\left(\begin{array}{c}
b_1 \\ b_2
\end{array}\right)=\cM_{12}(k_\star)\left(\begin{array}{c}
a_3 \\ a_4
\end{array}\right).
\end{equation}
Hereafter $0_{m,n}$ stands for a $m\times n $ zero matrix.

Thus there exist two possibilities for realization of the lasing. The first one corresponds to   the case when the determinant of matrix $\cM_{22}(\ks)$ is zero [i.e., $\Delta(\ks)=0$] but some of the matrix elements of $\cM_{22}(\ks)$ are non-zero. In this case the amplitudes of  left-incident waves $a_3$ and $a_4$ are determined by first of the equations (\ref{laser}).
Such a spectral singularity will be called   {\em weak}.
The second possibility corresponds to the case when all entries of the $2\times 2$ block $\cM_{22}$ are zero, i.e.,  $\cM_{22}(k_\star)=0_{2,2}$. This singularity will be termed   as {\em  strong}; it corresponds to the   arbitrary choice of the amplitudes $a_{3,4}$, including the cases when one of them is zero.

Similarly, the CPA corresponds to solutions 
\begin{eqnarray}
\label{cpa_solut}
\ba_{CPA}=(a_1,a_2,0,0)^T, \quad  \bb_{CPA}=(0,0,b_3,b_4)^T, 
\end{eqnarray}
i.e., to the conditions
\begin{equation}
\label{CPA}
\cM_{11}(k_\star)\left(\begin{array}{c}
a_1 \\ a_2
\end{array}\right)=0_{2,1},
\quad   
\left(\begin{array}{c}
b_3 \\ b_4
\end{array}\right)=\cM_{21}(k_\star)\left(\begin{array}{c}
a_1 \\ a_2
\end{array}\right),
\end{equation}
Such $k_\star>0$ is a time-reversed SS, which is either {\em weak}, when $\det\cM_{11}(k_\star)=0$, but $\cM_{11}(k_\star)\ne0_{2,2}$, or {\em strong}, when $\cM_{11}(k_\star)=0_{2,2}$.

In the case of a strong SS  
one has:
\begin{eqnarray}
\label{eq:partition}
\det M(k_\star)=-\det\cM_{12}(k_\star)\, \det\cM_{21}(k_\star).
\end{eqnarray}
This relation together with Eq.~(\ref{eq:detM1}) means that antigiagonal blocks $\cM_{ij}(k_\star)$ are invertible.  Using a similar argument,  in the case of a weak SS, one can establish that antidiagonal blocks $\cM_{ij}(k_\star)$ are nonzero matrices. However, their determinants in principle may acquire the zero value.
Let for a given weak singularity $\det\cM_{12}(k_\star)=0$. Then, at least one of the eigenvalues of  $\cM_{12}(k_\star)$ is zero. Let also $\lambda$ be the second eigenvalue (it can be zero, too). Then by the Schur decomposition, the non-zero matrix $\cM_{12}(k_\star)$ by a unitary transformation $Q$ can be reduced to the upper triangular matrix of the form 
\begin{eqnarray}
\label{Schur}
\cM_{12}(k_\star)= Q \left(\begin{array}{cc}
0 &\mu \\ 0 & \lambda
\end{array}\right)Q^{-1},
\end{eqnarray}
where $\mu$ is a complex number. Hence, $[\cM_{12}(k_\star)]^2=\lambda \cM_{12}(k_\star)$. Thus applying  $\cM_{12}(k_\star)$ to both sides of the second equation in (\ref{laser}) we obtain that $(b_1,b_2)^T$ is an eigenvector of $\cM_{12}(k_\star)$  corresponding to the eigenvalue $\lambda$. Similarly, if for some CPA solution one has $\det \cM_{21}(\ks)=0$, then    $(b_3,b_4)^T$ is an eigenvector of $\cM_{21}(k_\star)$.  

Summarizing, for any CPA ($j=1$) or lasing  ($j=2$) solution of the potential characterized by a transfer matrix with a non-invertible antidiagonal block $\cM_{ij}$ ($i\neq j$), the field from one side of the potential is an eigenvectors of the diagonal block $\cM_{jj}$ while the field from the opposite side of the potential is an eigenvector of the antidiagonal block.

\section{Self-dual spectral singularities}
\label{sec:SS}
 

The definitions and    properties described in the previous section did not assume the system to admit any particular symmetry. In this section, we use the odd $\PT$ symmetry (\ref{PT})  to obtain  additional information on the scattering properties. 

Since the Jost solutions are defined uniquely by their asymptotics, odd $\PT$ symmetry results in the relations
\begin{eqnarray*}
\label{eq:PTphi}
&&\PT\bphi_{11}(x,k)=i\bphi_{22}(x,k^*),
\\
&&\PT\bphi_{31}(x,k)=i\bphi_{42}(x,k^*),
\\
&&\PT\bphi_{21}(x,k)=-i\bphi_{12}(x,k^*),
\\
&&\PT\bphi_{41}(x,k)=-i\bphi_{32}(x,k^*).
\end{eqnarray*}
Using these formulas, one can establish the property:
\begin{eqnarray}
\label{inverseM}
M^{-1}(k^*)=  SM^{*}(k) S,  \quad \mbox{where} \quad  S=\sigma_0\otimes\sigma_2,
\end{eqnarray}
which in the explicit form means that
\begin{eqnarray}
\label{inverseM_expl}
M^{-1}(k^*)=   \left(
\begin{array}{cccc}
M_{22}^* & -M_{21}^* &M_{24}^* & -M_{23}^*
\\
-M_{12}^* & M_{11}^* &-M_{14}^* & M_{13}^*
\\
M_{42}^* & -M_{41}^* & M_{44}^* & -M_{43}^*
\\
-M_{32}^* & M_{31}^* &-M_{34}^* & M_{33}^*
\end{array}
\right),
\end{eqnarray} 
where all matrix elements in the right hand side are evaluated at  $k$ (not at $k^*$). 
 
 
Now we prove that any SS, either weak or strong, is self-dual, i.e., that any SS at the wavenumber $\ks$ is always accompanied by the time-reversed SS at the same wavenumber $\ks$.

As discussed above, for the strong SS with $\cM_{22}(\ks)=0_{2,2}$, the matrices $\cM_{12}$ and $\cM_{21}$ are invertible. Thus at $k=\ks$ we have  \cite{matrix}
\begin{eqnarray}
  M^{-1}=\left(
\begin{array}{cc}
 0_{2,2} & \cM_{21}^{-1}
\\
\cM_{12}^{-1} & -\cM_{12}^{-1}\cM_{11}\cM_{21}^{-1}
\end{array}
\right).
\end{eqnarray}
Comparing this expression with (\ref{inverseM}) [or (\ref{inverseM_expl})], we observe that the block $11$ of $M^{-1}$ has the form (below the blocks of $M^{-1}$ are denoted as $\tcM_{ij}$ with $i,j=1,2$)
\begin{eqnarray}
\label{tcM11}
\tcM_{11}=\left( 
\begin{array}{cc}
M_{22}^* &-M_{21}^* \\
- M_{12}^* &M_{11}^* 
\end{array}
\right)
\end{eqnarray}
Thus $\cM_{22}(k_\star)=0_{2,2}$ implies $\cM_{11}(k_\star)=0_{2,2}$. The converse can be proven in a similar way: if   $\cM_{11}(k_\star)=0_{2,2}$, then  $\cM_{22}(k_\star)=0_{2,2}$.  

Now we turn to weak singularities, i.e., assume that $\Delta(\ks)=\det \cM_{22}(\ks)=0$. 
Suppose that a weak SS is not self-dual, i.e., $\det \cM_{11}(\ks) \ne 0$. Since the entire transfer matrix $M(\ks)$ is invertible, we  conclude that there  exists an invertible matrix $\cC_1(k_\star)$,  defined as 
\begin{equation}
\label{C1}
\cC_1(\ks) = \cM_{22}(\ks) - \cM_{21}(\ks)[\cM_{11}(\ks)]^{-1}\cM_{12}(\ks), 
\end{equation}
and the block $22$ of the inverse matrix $M^{-1}(\ks)$ is equal to $[\cC_1(\ks)]^{-1}$
[see \cite{matrix} or formula (\ref{app:C1}) in  Appendix~\ref{app:B}]. On the other hand, as follows from (\ref{inverseM_expl}), the determinant of this block is equal to $\det \cM_{22}^*(\ks)$, and the latter is equal to zero. Hence the determinant of $[\cC_1(\ks)]^{-1}$ is zero, which contradicts to the invertibility of  $\cC_1(\ks)$. The contradiction can only be removed if we admit  that $\det \cM_{11}(\ks)=0$. Therefore,  $\det \cM_{22}(\ks)=0$ implies  $\det \cM_{11}(\ks)=0$. The converse statement, i.e., that $\det \cM_{11}(\ks)=0$ implies  $\det \cM_{22}(\ks)=0$, can be  proven in a similar way, but  employing the matrix
\begin{equation}
\label{C2}
\cC_2(\ks) = \cM_{11}(\ks) - \cM_{12}(\ks)[\cM_{22}(\ks)]^{-1}\cM_{21}(\ks) 
\end{equation}
with the formula for the inverse of the block matrix (\ref{app:C2}). 

Thus, we have proven that {\em both strong and weak spectral singularities are self-dual}. This result extends the known property of spectral singularities of (even) $\PT$-symmetric systems~\cite{Longhi} to the case of odd-$\PT$ symmetry.

Alternatively, the self-dual nature of any spectral singularity can be established using the property of $\PT$ symmetry. Indeed, from the asymptotic behavior of lasing and CPA solutions it follows that applying $\PT$ operator to a lasing solution with $k=\ks$ one obtains a CPA one with the same wavevector $\ks$ (and vice versa).  Thus lasing and CPA always take place simultaneously, i.e., at the same wavevector $\ks$.


Let us now assume spectral parameter $k$ to be complex-valued and consider the matrix functions $\cM_{ij}(k)$ defined in the complex plane. 
Let us also  assume that the matrix potential $\hat{U}(x)$ depends on one (or several) real parameters, say $v_j$ ($j=0,1...$), and that $k_\star>0$ is a (self-dual) weak spectral singularity for given values of the parameters, say for $v_{\star j}$. Then, subject to the variation of some of the parameters  $v_j$,
the self-dual spectral singularity $\ks$   either moves along real axis remaining self-dual or disappears. The latter case corresponds to the situation when  the zeros   of $\det\cM_{11}(k)$ and $\det\cM_{22}(k)$ split~\cite{HHK} and move to the complex plane. As a result, a   pair of complex-conjugate eigenvalues emerges in the spectrum of the Hamiltonian. Such a \textit{splitting of the self-dual spectral singularity} into a complex conjugate pair  is one of the possible scenarios of the $\PT$-symmetry breaking, different from the coalescence of the discrete eigenvalues at an exceptional point~\cite{KZ2017}. 

Like in   even-$\PT-$symmetric systems, 
the described splitting of a self-dual SS is constrained by the symmetry. Indeed, let  for some  complex $k_0$ we have   $\det\cM_{22}(k_0)=0$, but $\det\cM_{11}(k_0^*)\neq 0$. Then, as follows from  Appendix~\ref{app:B}, $\cC_1(k_0^*)$  is invertible, and the  lower diagonal block of $M^{-1}(k_0^*)$ is equal to  $[\cC_1(k_0^*)]^{-1}$. However, from the explicit expression  (\ref{inverseM_expl}) it follows that the determinant of the latter block is equal to $[\det\cM_{22}(k_0)]^*$ which is zero.
This contradicts to the invertibility of $\cC_1(k_0^*)$. The contradiction can be lifted only if we admit that $\det\cM_{11}(k_0^*) = 0$.
 
Similarly, one can prove that if $\det\cM_{11}(k_0)=0$, then $\det\cM_{22}(k_0^*)=0$. In other words, {\em in a $\PT$-symmetric system, roots of $\det\cM_{11}(k)$ and $\det\cM_{22}(k)$ are either self-dual or complex conjugate.}

\section{An example} 
\label{sec:example} 
 
 \subsection{Simplified model}
 
Now we turn to a specific example, where in Eq.~(\ref{Hamiltonian}) it is set: $U_0(x)=U_1^*(x)=V_0(x)+iV_1(x)$, and
\begin{equation}
\label{model1}
V_{0,1}(x)=V_{0,1}(-x), \qquad
    V_2(x)=-V_2(-x),
\end{equation}
where $V_{0}(x)$, $V_{1}(x)$, and $V_{2}(x)$ are real-valued functions.
This leads to the following, still quite general, scattering problem  
 \begin{eqnarray}
\label{main_stationaty}
 \begin{array}{l}
 -\psi_{1,xx}+[V_0(x)+iV_1(x)]\psi_1+V_2(x)\psi_2=k^2\psi_1,
 \\[2mm]
 -\psi_{2,xx}+[V_0(x)-iV_1(x)]\psi_2+V_2(x)\psi_1=k^2\psi_2.
  \end{array}
\end{eqnarray}
According to (\ref{eq:limits}), we require $
\lim_{x\to\pm\infty}V_{0,1,2}(x)=0.$ 

For the particular form of Hamiltonian (\ref{Hamiltonian}) corresponding to model (\ref{main_stationaty}) one can identify additional linear, $\sigma_3\p$, and antilinear, $\sigma_1 \K$, symmetries: 
\begin{eqnarray}
[H,\sigma_3 \p] =[H,\sigma_1 \K]=0_{2,2}.
\end{eqnarray} 
These symmetries lead to additional relations among the blocks of the transfer matrix $M(k)$, which means that  only the upper blocks  $\cM_{11}$ and $\cM_{12}$  are    sufficient to know the entire transfer matrix (or its inverse).  Indeed $\sigma_1\K$ symmetry means that for real $k$ the transfer matrix can be represented as
\begin{eqnarray}
\label{eq:s1K}
M=\left(\begin{array}{cc}
\cM_{11} & \cM_{12}
\\
\sigma_1\cM_{12}^*\sigma_1 & \sigma_1\cM_{11}^*\sigma_1
\end{array}\right).
\end{eqnarray} 
Using the symmetry $\sigma_3\p$ and the expression for the inverse transfer matrix (\ref{inverseM_expl}),  one obtains  
\begin{eqnarray}
\label{eq:s3P}
M^{-1}=\left(\begin{array}{cc}
\sigma_2\cM_{11}^*\sigma_2 & \sigma_2\cM_{12}^*\sigma_2
\\
\sigma_3\cM_{12}\sigma_3 & \sigma_3\cM_{11}\sigma_3
\end{array}\right).
\end{eqnarray} 

Furthermore, analyzing the 11 and 12 blocks of the identity   $MM^{-1}=I$ (where $I$ is $4\times 4$ identity matrix), written in terms of (\ref{eq:s1K}) and (\ref{eq:s3P}), one obtains the matrix relations connecting blocks $\cM_{11}$ and $\cM_{12}$ expressed in terms of the matrices: $\cN_{1}=\cM_{11}\sigma_2$ and $\cN_{2}=\cM_{12}\sigma_3$:
\begin{eqnarray}
\label{relat1}
\cN_{1}\cN_{2}^*=\cN_{2}\cN_{1}, \quad \cN_{2}^2-\cN_{1}\cN_{1}^* =\sigma_0.
\end{eqnarray}
Let $\alpha_{1,2}(k)$ be the two eigenvalues of the matrix $\cN_{2}(k)$. From the second equation in (\ref{relat1}), we have that 
\begin{equation}
|\det\cM_{11}|^2=\det (\sigma_0-\cN_2^2 )=(1-\alpha_1^2) (1-\alpha_2^2).
\end{equation}
Thus,   any self-dual spectral singularity (i.e., the moment when $\det\cM_{11}(\ks)=\det\cM_{22}(\ks)=0$)  takes place when   at least one of the eigenvalues of $\cM_{12}\sigma_3$  is equal to  either $+1$ or $-1$. 

If $k$ is not a spectral singularity, and thus $\cN_{1}(k)$ is invertible, from the first equation in (\ref{relat1}) we obtain that $\cN_{2}$ and $\cN_{2}^*$ are similar, i.e., they share the same eigenvalues which are either  a complex conjugate pair or both real. In the latter case they must satisfy   $\alpha_{1,2}>1$ or $\alpha_{1,2}<1$. Thus, if $k$ is not a spectral singularity, then $\det\cM_{12}(k)=-\det\cN_2(k)$ is real.


Let now $k=k_\star$. Consider a CPA solution.  From the relation $\ba=M^{-1}\bb$ [see (\ref{bMa})] and from the explicit expressions (\ref{eq:s1K}) and (\ref{eq:s3P}) one can deduce that   
 \begin{equation}
 	\label{eq:aux1}
 \cN^2\tilde{\bb}=\tilde{\bb}, \quad\mbox{where}\quad 	\cN=\sigma_3\cM_{12}^*(k_\star), \quad \tilde{\bb}=\sigma_2 
 	\left(\begin{array}{c}
 		b_3 \\ b_4
 	\end{array}\right). 
 \end{equation}
Thus the CPA solution incident from $+\infty$ (up to the normalization amplitude) can be computed directly using the antidiagonal blocks of the transfer matrix at the SS. Result (\ref{eq:aux1}) also means that at least one of the eigenvalues of $\cN$ is either $+1$ or $-1$, and hence $\cM_{12}(k_\star)$ has at least one nonzero eigenvalue. If additionally $\det \cM_{12}(k_\star)=0$, i.e., one of the eigenvalues of the block $\cM_{12}(k_\star)$ is zero, then $\det\cN=0$ and we conclude that $|$Tr$\,\cN|=|$Tr$\,\cM_{12}(k_\star)|= 1$. Thus the second eigenvalue  if $\cM_{12}(k_\star)$ must be either $1$ or $-1$. This also leaves us with only one of two possibilities discussed above in Sec.~\ref{sec:CPA-lasing}: if $\det\cM_{12}(k_\star)=0$, then $(b_3,b_4)^T$ is an eigenvector of $\cM_{21}(k_\star)$ corresponding to either $1$  or $-1$ eigenvalue (which corroborates with the respective conclusion made above).

\subsection{Physical models}
\label{sec:phyics}

Model (\ref{main_stationaty}) describes various physical settings. Below we present two of them.

\subsubsection{Non-Hermitian odd-$\PT$-symmetric coupler}
\label{sec:antiPT}

Optical couplers allow for modeling almost any type of symmetries (see e.g.~\cite{Konotop}). In particular, couplers for monochromatic waves allow for simulating odd-$\PT$ symmetry~\cite{ZK2018}. Thus it is natural to model stationary scattering by odd-$\PT$-symmetric potentials using a dispersive coupler.

\begin{figure}
	\centering
	\includegraphics[width=\columnwidth]{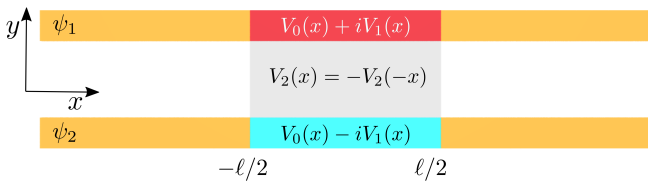}
	\caption{Schematic illustration  of two waveguides  which are locally coupled through an antisymmetric medium  characterized by $V_2(x)$. In the domain of coupling, the waveguides  are doped by active impurities resulting in gain, $+iV_1(x)$, and loss, $-iV_1(x)$. A paraxial beam propagates along $z$-axis directed toward the reader.}
	\label{fig:one}
\end{figure}

Indeed, consider two transparent planar waveguides located in the plane $(x,z)$ and separated along $y$-axis as illustrated in Fig.~\ref{fig:one}. In each waveguide quasi-parallel to the $z-$axis there propagates a monochromatic paraxial beam. The fields of the beams are polarized in the planes of the waveguides. They are denoted by $\Psi_{1}$ and $\Psi_{2}$ for the first and second waveguides, respectively. The separation of the waveguides is sufficiently strong to allow one to neglect overlapping of the beams. In the domain $x\in[-\ell/2,\ell/2]$ the space between the waveguides is filled with an antisymmetric 
medium, characterized by the optical potential $V_2(x)=-V_2(-x)$, 
which enhances the evanescent coupling. Also in the same interval, $x\in[-\ell/2,\ell/2]$, the waveguide cores are doped by impurities making one of the waveguides active and another one absorbing. Propagation of the paraxial beams  $\Psi_{1,2}=e^{i\beta z}\psi_{1,2}(x)$, where $\beta$ is the propagation constant, in such a coupler is described by Eqs.~(\ref{main_stationaty}) with
\begin{eqnarray}
\label{V1_odd}
V_{0,1}(x)&=&\left\{\begin{array}{ll}
v_{0,1} & |x|< \ell/2 
\\
0 &|x|> \ell/2
\end{array}\right., 
\\
\label{V2_odd}
V_{2}(x)&=&\left\{\begin{array}{ll}
-v_{2} &  x\in(-\ell/2,0) 
\\
v_{2} &  x\in(0, \ell/2)
\\
0 &|x|> \ell/2
\end{array}\right. ,
\end{eqnarray}
where the amplitudes $v_{0,1,2}$ are real and $k^2=-\beta$. 

\subsubsection{Scattering of condensates of two-level atoms}

 System (\ref{main_stationaty}) can be also considered  as coupled stationary Gross-Pitaevskii equations governing a binary mixture of quasi-one-dimensional Bose-Einstein condensates of hyperfine states of two-level atoms, at zero nonlinearity. The states are described by macroscopic wavefunctions $\psi_{1,2}$ and the condensate densities are assumed to be low enough to neglect two-body interactions. It is assumed also, that in the vicinity of $x=0$ atoms are loaded in one hyperfine state and are removed from the other state. The coupling is implemented by a synthetic magnetic field, which has opposite directions at $x<0$ and $x>0$ originating Zeeman splitting, is described by $V_2(x)$ which is anti-symmetric with respect to $x$. In this case the potentials $V_{0,1,2}$ can be also modeled by Gaussian shapes. The effect of the interatomic interactions can be accounted by adding the respective nonlinear terms. 


\subsection{Spectral singularities of coupled waveguides}

Below we concentrate on the model of optical coupler  described by (\ref{V1_odd}) and (\ref{V2_odd}). 

It has been shown above that any spectral singularity is self-dual in an odd-$\PT$-symmetric system. Existence of a strong SSs requires simultaneous solution of four complex equations assuring zero values of the entries of $\cM_{jj}$, which seems to be hardly possible in our system where  only four real-valued parameters are available: $v_{0,1,2}$ and $\ell$. Therefore, we look for weak SSs. To this end, it is sufficient to satisfy only one equation $\det \cM_{11}(\ks) = 0$, and  the equation $\det\cM_{22}(\ks)=0$ will be satisfied automatically. Since $\det \cM_{11}(k)$ is   a complex-valued function of a real parameter $k$ (recall we are interested only in real roots $k_\star$), in order to find a weak SS one needs at least one real-valued control parameter. As such  a parameter we will use the ``strength of the non-Hermiticity'', i.e., $v_1$ characterizing gain and loss (holding $v_{0,2}$ fixed). Once a SS is found, by changing a second parameter, say $v_2$, one can construct a curve which shows the position of spectral singularity in a parametric space $(v_1, v_2)$, with $v_0$ remaining fixed constant (all points of such a curve, however, correspond to different $k_\star$). 

Without loss of generality, in what follows we will restrict our attention to $v_1\geq 0$ and $v_2\geq 0$. At the same time, the properties of the system  depend significantly on the sign of the parameter $v_0$.  The cases of positive and negative values of $v_0$ are considered separately.

\subsubsection{Potential barrier, $v_0>0$}

\begin{figure}
	\centering
		\includegraphics[width=1.00\columnwidth]{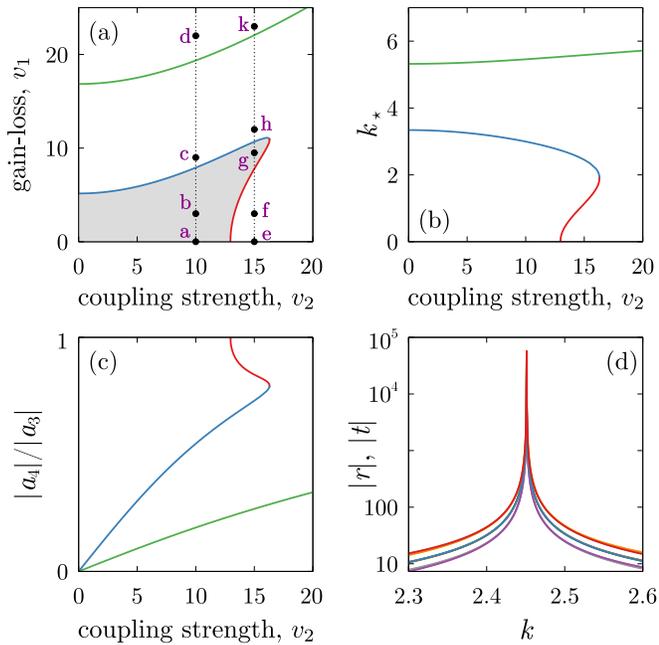}
	\caption{Scattering by the odd-$\PT$-symmetric potential barrier with $v_0=8$ and $\ell=1$. (a) Values of $v_1$ and $v_2$ which correspond to  weak SSs are plotted as curves in the $(v_1, v_2)$ plane. The lower (red and blue) lines separate the domains of
		unbroken (shaded area) and broken (white area) $\PT$-symmetric phases. The black dots labeled a-k correspond to panels (a)--(k) in Fig.~\ref{fig:obeig}. (b) Weak self-dual SSs, $k_\star$  as   functions of the coupling strength $v_2$. (c) Ratio $|a_4/a_3|$ between the amplitudes of the two polarizations for the laser solution.  Different colors in (a,b,c)  are used to visualize the correspondence between the  curves in different panels.    
		(d) Magnitudes of reflection and transition coefficients  $r^{R,L}_{\alpha\beta}$ and $t^{r,l}_{\alpha \beta}$ defined in Appendix~\ref{app:A} as functions of the wavenumber $k$ driven trough the self-dual SS for $v_1\approx10.685$,  $v_2=15$.
	}
	\label{fig:obss} 
\end{figure}

\begin{figure}
	\centering
		\includegraphics[width=1.00\columnwidth]{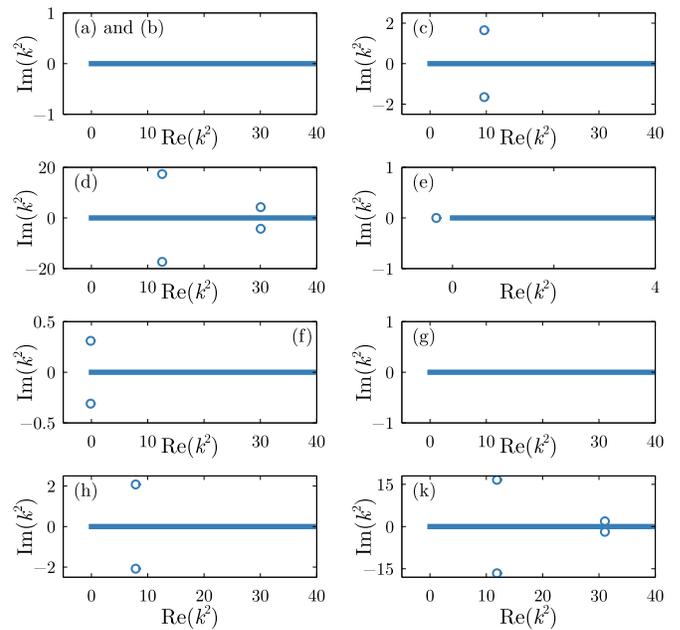}
	\caption{Real and imaginary parts of eigenvalues for   different combinations of parameters labeled with black dots in Fig.~\ref{fig:obss}(a). Thick line shows the continuous spectrum, and circles correspond to isolated eigenvalues. }
	\label{fig:obeig} 
\end{figure}

Resulting dependencies for a representative value $v_0=8$  are shown in Fig.~\ref{fig:obss}(a,b). 
In order to get insight into the main features of the scattering problem, it is instructive to trace its  behavior moving   along the vertical axis in Fig.~\ref{fig:obss}(a), which corresponds to the increase of  the non-Hermicity parameter $v_1$ departing from  the Hermitian limit $v_1=0$. In the latter limit     the fields $\psi_1\pm\psi_2$ are decoupled and each is subject to effective real potential $V_0(x)\pm V_2(x)$. The spectrum of the Hermitian system is obviously real. If the coupling strength $v_2$ is small enough [point a in Fig.~\ref{fig:obss}(a)], then  at $v_1=0$ the spectrum is purely continuous and fills the semiaxis $k^2>0$. For sufficiently large coupling $v_2$ [point e in Fig.~\ref{fig:obss}(a)], the spectrum of the Hermitian system additionally contains a degenerate isolated  eigenvalue   with $k^2<0$ [Fig.~\ref{fig:obeig}(e); recall that in a system with  odd $\PT$ symmetry any   real eigenvalue corresponds to at least two linearly independent eigenvectors, which is a direct consequence of the property $(\PT)^2=-1$]. 

Qualitatively different spectra in the Hermitian limit for small and large values of the coupling coefficient $v_2$ result in  essentially different behaviors of the system  subject to  the increase of the non-Hermiticity strength $v_1$. If at $v_1=0$ the spectrum is purely real and continuous [point a in Fig.~\ref{fig:obss}(a)], then it  remains so for sufficiently small but nonzero  $v_1$ [point b in Fig.~\ref{fig:obss}(a)]. Further increase  of $v_1$  results in  the $\PT$-symmetry breaking through the splitting of the self-dual SS. In   Fig.~\ref{fig:obss}(a) it occurs when the vertical dashed line intersects the blue curve between points b and c. As a result, a single complex conjugate pair   emerges from the continuous spectrum, see Fig.~\ref{fig:obeig}(c).  This scenario of $\PT-$symmetry breaking  through the {\em splitting of a self-dual SS}  was described in~\cite{KZ2017} for a scalar scattering problem. It represents the most typical mechanism of $\PT$-symmetry breaking  in systems without the discrete spectrum. Further increase of $v_1$ drives the system trough a new spectral singularity which occurs when the vertical dashed line intersects the green curve between points  c and d in Fig.~\ref{fig:obss}(a). Respectively, for sufficiently large values of $v_1$, the spectrum contains two complex conjugate pairs as shown in  Fig.~\ref{fig:obeig}(d).  It is natural to expect that the further increase of $v_1$ may result in new spectral singularities and new complex conjugate pairs emerging from the continuous spectrum.

Behavior of the system becomes more complicated for larger values of the coupling strength $v_2$ [see points e, f, g, h, and k in Fig.~\ref{fig:obss}(a) and the corresponding panels in Fig.~\ref{fig:obeig}]. As $v_1$ departs from zero, the real double isolated eigenvalue immediately splits into a complex conjugate pair [point f in Fig.~\ref{fig:obeig}(a)]. This is the thresholdless  $\PT$-symmetry breaking. However, as $v_1$ increases further,  the complex conjugate pair returns to the real axis -- this   corresponds to  the SS at the intersection between the vertical dashed line and the red $v_2(v_1)$ curve  between points f and g  in Fig.~\ref{fig:obss}(a). Thus the increase of the non-Hermiticity parameter through the self-dual spectral singularity drives the system from  $\PT$-broken [point f in Fig.~\ref{fig:obss}(a) and Fig.~\ref{fig:obeig}(f)] to unbroken phase [point g in Fig.~\ref{fig:obss}(a) and Fig.~\ref{fig:obeig}(g)]. Hence  we encounter a rather interesting situation when \textit{the increase of the non-Hermiticity strength leads to the restoration of unbroken $\PT$ symmetry trough the coalescence of two isolated complex conjugate eigenvalues   at a self-dual spectral singularity}. This  effect is opposite to the $\PT$-symmetry breaking through the splitting of a self-dual SS.
Further increase of $v_1$ triggers  a new self-dual SS [at the intersection between the dashed line and the blue   curve between points g and h, as shown in Fig.~\ref{fig:obeig}(h)], which results in the $\PT$-symmetry breaking with  a new complex conjugate pair of eigenvalues emerging from  
the  continuous spectrum. 
Next, the increase of $v_1$ from point h to point k results in a new spectral singularity, where the second complex conjugate  pair emerges  [see Fig.~\ref{fig:obeig}(k)]. 

To give physical interpretation for the splitting of self-dual SS,  let us recall the relations (\ref{laser}) and (\ref{CPA}). If Im$\,k_0<0$, where $k_0$ is a zero of $\det\cM_{11}(k_0)=0$ and respectively  $\det\cM_{22}(k_0^*)=0$, the system allows for a solution $\bpsi_{-}$ that transforms into the CPA solution at Im$k_0\to -0$  and outside the potential, i.e., at $|x|>\ell/2$, can be written down as
\begin{equation}
\label{BIC_CPA}
\begin{array}{ll}
\displaystyle{\bpsi_{-}=e^{ik_0x}[  M_{12}(k_0)\up -M_{11}(k_0)\dn}] , & \displaystyle{x<-\frac{\ell}{2}},
\\
\displaystyle{\bpsi_{-}=e^{-ik_0x}[  M_{12}(k_0)\up +M_{11}(k_0)\dn]}, & \displaystyle{x>\frac{\ell}{2}}.
 \end{array}
\end{equation}
as well as for a solution $\bpsi_{+}$ which transforms in to the lasing solution in the limit Im$k_0\to -0$:
\begin{equation}
\label{BIC_laser}
\begin{array}{ll}
\displaystyle{\bpsi_{+}=e^{-ik_0^*x}[ M_{34}(k_0)\up -M_{33}(k_0)\dn]}, & \displaystyle{x<-\frac{\ell}{2}},
\\
\displaystyle{\bpsi_{+}=e^{ik_0^*x}[ M_{34}(k_0)\up + M_{33}(k_0)\dn]}, & \displaystyle{x>\frac{\ell}{2}}.
\end{array}
\end{equation}
These are spatially localized solutions with the complex energy $E=k_0^2$ [Eq.~(\ref{BIC_CPA})] and $E=[k_0^*]^2$ [Eq.~(\ref{BIC_laser})]. In optical applications, where $-E=\beta$ (see discussion in Sec.~\ref{sec:antiPT}) is the propagation constant, these solutions represent propagating beams whose intensity decays in the case (\ref{BIC_CPA}) and grows in the case (\ref{BIC_laser}), along propagation (i.e. along $z$-axis). Therefore they are loosely referred to as bound states in continuum (BICs)~\cite{BIC_optics,KZ2017} (should not be confused with BIC in the conventional mathematical sense, which are spatially localized eigenstates with real energy embedded in the continuous spectrum~\cite{BIC_true}).

CPA and lasing solutions 
are  superpositions of two polarizations.  The amplitudes of different polarizations are related to each other by the formula  
\begin{eqnarray}
\label{amp_cpa}
-\frac{a_1}{a_2}=\frac{b_3}{b_4}=\frac{M_{12}(k_\star)}{M_{11}(k_\star)}
\end{eqnarray}
for the CPA solution, and by the formula 
\begin{eqnarray}
\label{amp_las}
-\frac{a_3}{a_4}=\frac{b_1}{b_2}=\frac{M_{11}^*(k_\star)}{M_{12}^*(k_\star)}  
\end{eqnarray}
for the lasing solution. Numerically we obtained that the amplitude ratio between up and down polarization is always below unity $|a_4/a_3| = |b_2/b_1|<1$ [shown in Fig.~\ref{fig:obss}(c)]. Physically this is an expected result: it means that the laser solution is dominantly concentrated in the mode $\up$, i.e., in the  waveguide with gain. Conversely, for CPA solution the spin-down component dominates, which means that the respective solutions are concentrated in the lossy waveguide. Interestingly, the relation between the amplitudes of polarizations, have different slopes for SS belonging to different branches of the SS. Comparing Fig.~\ref{fig:obss}(c) with Figs.~\ref{fig:obss}(a) and~(b) we observe that the SS emerging    from the coalescing  isolated   complex eigenvalues (the red lines) has larger relation $|a_4/a_3|$, as compared with SS emerging from the continuous spectrum (blue and green lines).

Finally, computing the left and right reflection and transmission coefficients (see Appendix~\ref{app:A}), we observe that all of them diverge simultaneously as the system is driven through the self-dual SS, as this is illustrated in Fig.~\ref{fig:obss}(d) where we plot amplitudes of all sixteen coefficients $r^{R,L}_{\alpha\beta}$ and $t^{R,L}_{\alpha \beta}$ as functions of $k$ with all other parameters kept fixed and corresponding to a spectral singularity at $\ks \approx  2.451$.   Notice that some of the scattering coefficients have equal amplitudes; hence the   number of curves visible  in Fig.~\ref{fig:obss}(d) is less than sixteen.

\subsubsection{Potential well, $v_0<0$}

\begin{figure}
	\centering
	\includegraphics[width=1.00\columnwidth]{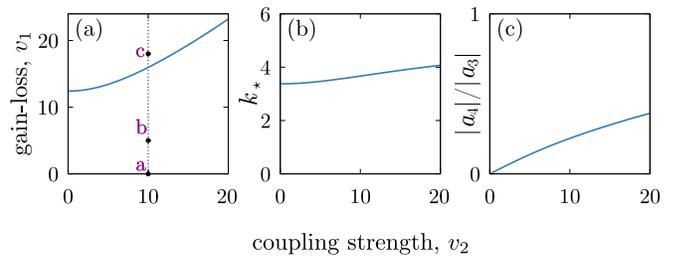}
	\caption{Scattering on the odd-$\PT$-symmetric potential well with $v_0=-3$ and $\ell=1$. (a) Values of $v_1$ and $v_2$ that correspond to weak SSs are plotted as curves in the $(v_1, v_2)$ plane. (b) Dependence of $k_\star$ on the coupling strength  $v_2$. (c) The ratio $|a_4/a_3|$ between amplitudes of the two polarizations of the lasing solution. Black dots labeled a-c correspond to panels (a)--(c) in Fig.~\ref{fig:oweig}. Here $\PT$ symmetry is  broken for any $v_1>0$ with one or more complex conjugate pairs in the spectrum.}
	\label{fig:owss} 
\end{figure}

\begin{figure}
	\centering
		\includegraphics[width=1.00\columnwidth]{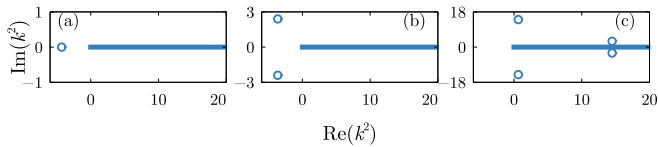}
	\caption{Real and imaginary parts of eigenvalues for   different combinations of parameters labeled with black dots in Fig.~\ref{fig:owss}(a). Thick line shows the continuous spectrum, and circles correspond to isolated eigenvalues. }
	\label{fig:oweig} 
\end{figure}

The behavior of the system is significantly different for the scattering by a potential well, which corresponds to negative values of $v_0$. Choosing as a representative example $v_0=-3$, we illustrate  the spectral singularities and the corresponding  eigenvalue diagrams in Fig.~\ref{fig:owss} and Fig.~\ref{fig:oweig}. Starting from the Hermitian limit $v_1=0$,  we observe that for any value of the coupling strength $v_2$ the Hermitian system contains an isolated double eigenvalue, see point  a in Fig.~\ref{fig:owss}(a) and the corresponding spectrum shown in  Fig.~\ref{fig:oweig}(a). The nonzero non-Hermiticity parameter $v_1$, even infinitesimal, immediately results in the splitting of the double eigenvalue in a complex  conjugate pair, see point b in Fig.~\ref{fig:owss}(a) and the spectrum in Fig.~\ref{fig:oweig}(b). Thus the $\PT$-symmetry breaking in this case is thresholdless. Further increase of $v_1$ drives the system through a self-dual SS, which results in the emergence of a new complex conjugate pair [point c Fig.~\ref{fig:owss}(a) and Fig.~\ref{fig:oweig}(c)]. Thus, in the scattering by the potential well, the splitting of self-dual SS into a complex conjugate pair occurs, but it does not represent the boundary between the unbroken and broken $\PT$ symmetry, the latter being broken already for any nonzero non-Hermiticity strength $v_1$.

Additionally, we note that the dependence $v_1(v_2)$ corresponding to the spectral singularities increases monotonically  [Fig.~\ref{fig:oweig}(a)], the wavenumber $\ks$ depends weakly on the coupling strength $v_2$  [Fig.~\ref{fig:oweig}(b)], and the ratio between the amplitudes of two polarizations of the lasing solution [and respectively of the CPA solution, as it follows from (\ref{amp_cpa}) and (\ref{amp_las})] increases   with $v_2$, too  [Fig.~\ref{fig:oweig}(c)].

\subsection{Comparison with the even-$\PT$-symmetric coupling}
\label{subsec:symmetric}

To highlight the peculiarities of the scattering by odd-$\PT$-symmetric potentials, we complement the results collected above  with the analysis of a similar system, but coupled by a homogeneous (symmetric) medium. More specifically we consider   the model (\ref{model1})-(\ref{main_stationaty}), but with {\it even} function $V_2(x)$ in the form [cf.~(\ref{V2_odd})]
\begin{eqnarray}
V_{2}(x)=\left\{\begin{array}{ll}
v_{2} &  x\in(-\ell/2,\ell/2), \\
0 &|x|> \ell/2
\end{array}\right. .
\end{eqnarray}
Notice that this modification of the model affects only the total symmetry, but does not affect distribution of gain and losses. The resulting system is invariant under the conventional (even) parity-time reversal with the parity operator being $\p =\sigma_1$ and the bosonic time reversal $\T=\K$. At the same time, the new system does not respect the odd-$\PT$ symmetry introduced above.

\subsubsection{Potential barrier, $v_0>0$}

\begin{figure}
	\centering
		\includegraphics[width=1.00\columnwidth]{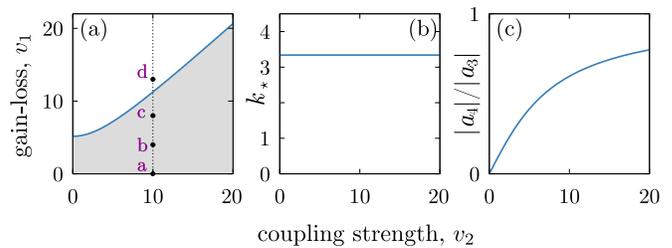}%
	\caption{ Scattering by the even-$\PT$-symmetric potential barrier with $v_0=8$ and $\ell=1$. 
		(a) Values of $v_1$ and $v_2$ which correspond to  weak SSs are plotted as curves in the $(v_1, v_2)$ plane and separate unbroken $\PT$-symmetric phase (shaded domain) and broken one (white domain). The black dots labeled a-c correspond to panels (a)--(c) in Fig.~\ref{fig:ebeig}. (b) The weak SS {\it vs} coupling strength. (c) The ratio $|a_4/a_3|$ between amplitudes of the two polarizations of the laser solution.      
	}
	\label{fig:ebss} 
\end{figure}

\begin{figure}
	\centering
	\includegraphics[width=1.00\columnwidth]{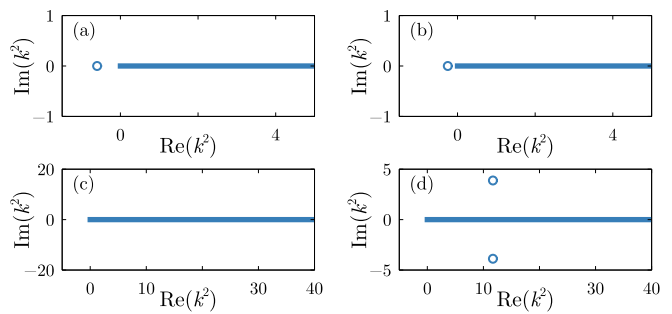}
	\caption{Real and imaginary parts of eigenvalues for   different combinations of parameters labeled with black dots in Fig.~\ref{fig:ebss}(a). Thick line shows the continuous spectrum, and circles correspond to isolated eigenvalues.}
	\label{fig:ebeig} 
\end{figure}

The results are summarized in Figs.~\ref{fig:ebss} and \ref{fig:ebeig}. Starting from the Hermitian limit, we observe that the spectrum is purely real and continuous for small values of the coupling $v_2$. However, for larger values of $v_2$,  the spectrum does contain one or more isolated eigenvalues.  In contrast to the odd-$\PT$-case, the isolated eigenvalues are generically simple. For the sake of illustration,  we choose coupling strength $v_2=10$, which corresponds to a single isolated eigenvalue in the Hermitian limit  [point~a in Fig.~\ref{fig:ebss}(a) and Fig.~\ref{fig:ebeig}(a)]. As $v_1$ departs from zero, the isolated eigenvalue remains real and moves towards the edge of the continuous spectrum [point~b in Fig.~\ref{fig:ebss}(a) and Fig.~\ref{fig:ebeig}(b)] and eventually merges  with the continuous spectrum. At this instant the spectrum becomes purely real and continuous  [point~c in Fig.~\ref{fig:ebss}(a) and Fig.~\ref{fig:ebeig}(c)]. Further increase   $v_1$ from point~c to point~d, leads to a self-dual spectral singularity which results to the $\PT$-symmetry breaking with a pair of complex conjugate eigenvalues emergin from the continuous   spectrum [point~d in Fig.~\ref{fig:ebss}(a) and Fig.~\ref{fig:ebeig}(d)]. 

The described behavior of the discrete eigenvalues and SSs for the even-$\PT$-symmetric coupler features several distinctive differences from those for the odd-$\PT$-symmetric coupler, described above.
First, in the case of even $\PT$ symmetry  we observe that dependence $v_1(v_2)$ (which separates unbroken and broken phases) is monotonic, see Fig.~\ref{fig:ebss}(a).  This is expectable, because the system is locally (i.e., at each given $x$) is $\PT$ symmetric and thus the $\PT$-phase transition for larger values of coupling $v_2$ is expected at larger gain and losses  $v_1$. In the case of odd-$\PT$-symmetry [Fig.~\ref{fig:obss}(a)] two new, somehow opposite, effects appear. At any value of gain and loss, i.e., of $v_1$, the increase of the coupling, $v_2$ results in $\PT$-symmetry breaking. Moreover, in some intervals of the coupling constant values, there exist two different values of gain-and-loss coefficient $v_1$ for which spectral singularities can be found. Thus for the coupling $v_2$ in this interval, the increase of the non-Hermiticity parameter $v_1$ can stabilize the odd-$\PT$-symmetric system, in a sharp contrast to destabilizing effect of growing $v_1$ (at fixed $v_2$) in the case of even-$\PT$ symmetry.  
Respectively, in the case of the odd-$\PT$-symmetry for a given $v_1$ 
lying in certain   specific intervals it is possible to obtain two spectral singularities, unlike this happens in the case of even-$\PT$ symmetry.
Also we observe, that for the even coupling $k_\star$ does not depend on $v_2$, and   the ratio between $|a_4|$ and $|a_3|$ computed for the laser solution grows monotonously  and slowly approaches unity.

When there are more than one eigenvalues in the Hermitian limit ($v_1=0$), the increase of $v_1$ leads to their successive immersion in the continuous spectrum, such  that after the last discrete eigenvalue is merged with the continuum, the spectrum remains purely continuous and real until the SS is formed. That is why the splitting of the self-dual SS always represents the boundary between unbroken and broken $\PT$ symmetry [i.e., between shaded and white areas in Fig.~\ref{fig:ebss}(a)].

\subsubsection{Potential well, $v_0<0$}

\begin{figure}
	\centering
	\includegraphics[width=1.00\columnwidth]{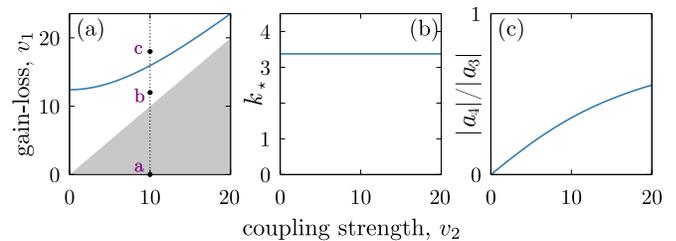}%
	\caption{Scattering by the even-$\PT$-symmetric potential well with   $v_0=-3$ and $\ell=1$.  (a) Values of $v_1$ and $v_2$ that correspond to  weak SSs are plotted as curves in the $(v_1, v_2)$ plane. Shaded and white domains correspond to the unbroken and broken $\PT$-symmetric phase, respectively. The black dots a, b, and c correspond to panels (a), (b) and (c) in Fig.~\ref{fig:eweig}. (b) The weak SS {\it vs.} the coupling strength $v_2$. (c) The ratio $|a_4/a_3|$ between amplitudes of the two polarizations of the laser solution.    
	}
	\label{fig:ewss}
\end{figure}

\begin{figure}
	\centering
	\includegraphics[width=1.00\columnwidth]{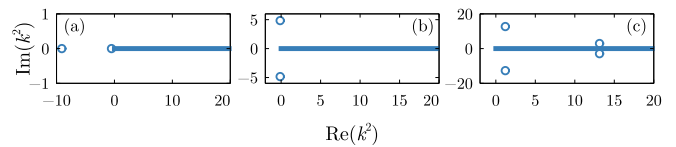}
	\caption{Real and imaginary parts of eigenvalues for   different combinations of parameters labeled with black dots in Fig.~\ref{fig:ewss}(a). Thick line shows the continuous spectrum, and circles correspond to isolated eigenvalues.}
	\label{fig:eweig}
\end{figure}

In the Hermitian limit $v_1=0$, the spectrum of the potential $\hU(x)$  with nonzero coupling $v_2>0$ contains the discrete part consisting of two or more  eigenvalues (depending on the depth of the well, $|v_0|$ and on  the coupling strength, $v_2$)  which  are generically simple. The increase of $v_1$ leads to  the  $\PT$-symmetry breaking trough the exceptional point scenario: at the threshold value of $v_1$, all real eigenvalues collide pairwise and simultaneously and then split in one (or several --- if the   well is deep enough) complex conjugate pairs. The representative diagrams with $v_0=-3$ (which corresponds to two real isolated eigenvalues in the Hermitian limit) are  presented in Figs.~\ref{fig:ewss} and \ref{fig:eweig}.
The phase transition corresponds  to the boundary between shaded and white domains in Fig.~\ref{fig:ewss}(a). It is also illustrated by the transition between points a and b in Fig.~\ref{fig:ewss}(a), as well as by transition between Fig.~\ref{fig:eweig}(a) and~(b). The value of non-Hermiticity $v_1$ where the  exceptional point transition takes place, is always below the value of $v_1$ corresponding to the self-dual SS, the latter shown by blue solid line in Fig.~\ref{fig:ewss}(a). 


\section{Discussion and conclusion}
\label{sec:conclusion}

In this work, we have developed formalism for the scattering of the two-component field by localized (matrix) potentials obeying  odd-$\PT$ symmetry. Being two-component, the reflected and transmitted fields have spinor character and can be characterized by two opposite polarizations. The resulting problem is described by a $4\times 4$ transfer matrix, which is naturally represented in the form of a $2\times2$ block matrix, where the blocks describe reflection, transmission, and transformation of waves with either of two polarizations.  The odd-$\PT$-symmetry relates two diagonal, as well as two anti-diagonal blocks. 

We identified two types of spectral singularities in such a system.  If the determinants of diagonal blocks vanish, one deals with a  weak singularity, for which lasing or CPA solutions are characterized by interrelated amplitudes of the polarizations incident or absorbed from one side of the potential. If the diagonal blocks are zero matrices,  one deals with a strong singularity, for which no constraints on the relations between the amplitudes are imposed. In any of these cases, spectral singularities are self-dual.

As an example we considered a simple odd-$\PT$-symmetric dispersive coupler with anti-symmetric coupling. The available parameters were not enough for obtaining strong spectral singularity, but allowed for a  detailed study of weak self-dual spectral singularities, as well as for the demonstration of the  odd-$\PT$-symmetry breaking through the splitting of   self-dual spectral singularities. 
We  have also found that in some cases the increase of the non-Hermiticity strength can lead to the restoration of the unbroken $\PT$-symmetry, which corresponds to the situation when two complex conjugate eigenvalues coalesce at the real axis and form of self-dual spectral singularity.   Apart from the phase transition through the spectral singularity, our simple odd-$\PT$-symmetric system features two other mechanisms of the transition from purely real to complex spectrum. One of these mechanisms is the well-studied coalescence of two isolated eigenvalues at the exceptional point, where the Hamiltonian becomes non-diagonalizable,  with the ensuing splitting of the multiple eigenvalue in a complex conjugate pair. Another encountered scenario corresponds to the splitting of a degenerate semi-simple eigenvalue (with algebraic and geometric multiplicities equal to two) into a complex conjugate pair (this situation  is distinctively different from the exceptional point scenario, because the Hamitonian with the semi-simple eigenvalue remains diagonalizable).

We have also compared properties of the odd-$\PT$-symmetric coupler with its even-$\PT$-symmetric counterpart and  reveals   significant qualitative differences in the scattering properties of the two systems. On the other hand, we can  also point  out some rather general features with hold for  either   even- and  odd-$\PT$- symmetric systems. In particular, we observe that in the parametric vicinity of each self-dual spectral singularity, the spectrum is either purely continuous or contains one or several complex conjugate pairs (a similar observation for the one-component scattering problem with the conventional even $\PT$ symmetry has been recently pointed out in \cite{Ahmed18}). Another intriguing observation is that  for all   considered cases complex conjugate pairs of   eigenvalues never coexist with real isolated eigenvalues.  Additionally, while for the scattering by the potential barrier  
the splitting of self-dual singularity can represent the boundary between unbroken and broken $\PT$ symmetry, i.e., the phase transition  occurs   through the splitting of the spectral singularity, 
for the scattering by the potential well the $\PT$-symmetry breaking   never occurs through the splitting of the self-dual spectral singularity. Instead, it is caused by a different mechanism (i.e., by the splitting of multiple isolated eigenvalues into complex conjugate pairs). As a result, for system with  the potential well, the splitting of a spectral singularity   results in the emergence of a new complex conjugate pair of complex eigenvalues in the already broken $\PT$-symmetric phase.

\begin{acknowledgments}
The research of D.A.Z. was supported by the Russian Science Foundation (Grant No. 17-11-01004) and by the government of the Russian Federation (Grant No. 08-08).
\end{acknowledgments}

\bigskip
\bigskip

\appendix

\section{Scattering data through the transfer matrix elements} 
  \label{app:A}  
  
Introduce 
\begin{equation*}
\Delta = M_{33}M_{44} - M_{34}M_{43}.
\end{equation*}
  
Left incidence:  
\begin{eqnarray*} 
\displaystyle{\ruu^L=\frac{M_{34}M_{41}-M_{31}M_{44}}{\Delta}}
 ,\,\,
\displaystyle{\rud^L=\frac{M_{31}M_{43}-M_{33}M_{41}}{\Delta}}
\\ 
\displaystyle{\rdu^L=\frac{M_{42}M_{34}-M_{32}M_{44}}{\Delta}}
,\,\,
\displaystyle{\rdd^L=\frac{M_{32}M_{43}-M_{42}M_{33}}{\Delta}},
\\
\displaystyle{\tuu^L=M_{11}+M_{13}\ruu^L+M_{14}\rud^L},
\\ 
\displaystyle{\tud^L=M_{21}+M_{23}\ruu^L+M_{24}\rud^L},
\\ 
\displaystyle{\tdu^L=M_{12}+M_{13}\rdu^L+M_{14}\rdd^L},
\\
\displaystyle{\tdd^L=M_{22}+M_{23}\rdu^L+M_{24}\rdd^L}.
\end{eqnarray*}

Right incidence:  
\begin{eqnarray*} 
\displaystyle{\ruu^R=\frac{M_{13}M_{44}-M_{14}M_{43}}{\Delta}}
,\,\,
\displaystyle{\rud^R=\frac{M_{23}M_{44}-M_{24}M_{43}}{\Delta}},
\\
\displaystyle{\rdu^R=\frac{M_{33}M_{14}-M_{13}M_{34}}{\Delta}}
,\,\, 
\displaystyle{\rdd^R=\frac{M_{24}M_{33}-M_{23}M_{34}}{\Delta}},
\\
\displaystyle{\tuu^R=\frac{M_{44}}{\Delta}}
,\,\,
\displaystyle{\tud^R=-\frac{M_{43}}{\Delta}}
,\,\,
\displaystyle{\tdu^R=-\frac{M_{34}}{\Delta}}
,\,\,
\displaystyle{\tdd^R=\frac{M_{33}}{\Delta}}.
\end{eqnarray*}  

\section{The inverse of the block matrix}
\label{app:B}

For the sake of convenience, here we present the explicit formulas for the inverse of a block matrix   \cite{inverse,matrix}.
Consider
\begin{eqnarray}
M=\left(\begin{array}{cc}
\cM_{11} & \cM_{12}
\\
\cM_{21} & \cM_{22}
\end{array}\right),
\end{eqnarray}
where $\cM_{ij}$ are $2\times 2$ matrices.

\begin{widetext}
	Assume that  $ \cM_{11}$ is nonsingular. Then $M$ is invertible if and only if   matrix $\cC_1$ defined as  $\cC_1=\cM_{22} -\cM_{21}\cM_{11}^{-1}\cM_{12}$, is invertible, and 
	\begin{eqnarray}
	\label{app:C1}
	M^{-1}=
	\left(
	\begin{array}{cc}
	\cM_{11}^{-1} + \cM_{11}^{-1}\cM_{12}\cC_1^{-1}\cM_{21}\cM_{11}^{-1} &  -\cM_{11}^{-1}\cM_{12}\cC_1^{-1}\\
	-\cC_1^{-1}\cM_{21}\cM_{11}^{-1}& \cC_1^{-1} 
	\end{array}
	\right).
	\end{eqnarray}

	Assume that  $ \cM_{22}$ is nonsingular. Then $M$ is invertible if and only if matrix $\cC_2$ defined as  $\cC_2=\cM_{11} -\cM_{12}\cM_{22}^{-1}\cM_{21}$, is invertible, and 
	\begin{eqnarray}
	\label{app:C2}
	M^{-1}= \left(
	\begin{array}{cc}
	\cC_2^{-1} & -\cC_2^{-1} \cM_{12}\cM_{22}^{-1}\\
	-\cM_{22}^{-1}\cM_{21}\cC_2^{-1}& 
	\cM_{22}^{-1} + \cM_{22}^{-1}\cM_{21}\cC_2^{-1}\cM_{12}\cM_{22}^{-1}
	\end{array}
	\right).
	\end{eqnarray}


%
\end{widetext}


\begin{thebibliography}{99}
	
	\bibitem{SS} M. A.  Naimark, Investigation of the spectrum and the expansion in eigenfunctions of a nonselfadjoint operator of the second order on a semi-axis, {\it Tr. Mosk. Mat. Obs.} {\bf 3}, 181--270 (1954);  J. Schwartz, Some non-selfadjoint operators, Comm. Pure Appl. Math. {\bf 13}, 609 (1960);  B. Vainberg, On the analytical properties of the resolvent for a certain class of operator-pencils, Math. USSR Sbornik, {\bf 6},
	241 (1968); G. S. Guseinov, On the concept of spectral singularities, Pramana - J. Phys. {\bf 73}, 587
	603 (2009).
		
	\bibitem{Mostafazadeh2009}  A. Mostafazadeh, 
	Spectral Singularities of Complex Scattering Potentials and Infinite Reflection and Transmission Coefficients at Real Energies, 	Phys. Rev. Lett. {\bf 102}, 220402 (2009). 
	
	\bibitem{ScarfII} Z. Ahmed, Zero width resonance (spectral singularity) in a complex PT-symmetric potential, J. Phys. A: Math. Theor. {\bf 42}, 472005 (2009).
	
	\bibitem{Longhi}   S. Longhi, $\PT$-symmetric laser absorber, Phys. Rev. A {\bf 82}, 031801 (2010).
	
	\bibitem{rectang} A. Mostafazadeh,
	Self-dual spectral singularities and coherent perfect absorbing lasers without $\PT$-symmetry, J. Phys. A: Math. Theor. {\bf 45}, 444024 (2012). 
	
	\bibitem{Stone} Y. D. Chong,  L.   Ge, H.   Cao, and A. D.  Stone, 
	Coherent Perfect Absorbers: Time-Reversed Lasers,
	{Phys. Rev. Lett.} {\bf 105}, 053901 (2010); W. Wan, Y. Chong, L. Ge, H. Noh, A. D. Stone, H. Cao, Time-reversed lasing and
	interferometric control of absorption. Science {\bf 331}, 889
	892 (2011).
	
	\bibitem{reviewCPA}  D. G. Baranov,  A.  Krasnok, T.   Shegai, A.  Al\'u,  and Y. D.   Chong, 
	Coherent perfect absorbers: linear control of light with light,
	{Nature Reviews Materials} {\bf 2},   17064 (2017).
	
		\bibitem{BenderBoet}  C. M. Bender and S. Boettcher, 
	Real Spectra in Non-Hermitian Hamiltonians Having $\PT$-Symmetry,
	Phys. Rev. Lett. {\bf 80} 5243 (1998); C. M. Bender, Making sense of non-Hermitian Hamiltonians, 
	Rep. Prog. Phys. {\bf 70}, 947
	(2007).
	
	\bibitem{Stone_selfdual} Y. D. Chong, L. Ge, and A. D. Stone, 
	PT-Symmetry Breaking and Laser-Absorber Modes in Optical Scattering Systems,
	 Phys. Rev. Lett. {\bf 106},
	093902 	(2011).
	
	\bibitem{KZ2017} V. V. Konotop and D. A. Zezyulin, 
	Phase transition through the splitting of self-dual
	spectral singularity in optical potentials,
	Opt. Lett. \textbf{42}, 5206 (2017).
	
	\bibitem{Mostafazadeh2009JPA} A. Mostafazadeh and H. Mehri-Dehnavi, Spectral singularities, biorthonormal systems and a two-parameter family of complex point interactions, J. Phys. A: Math. Theor. {\bf 42},  125303 (2009). 
	
	\bibitem{Kato} T. Kato, Perturbation Theory for Linear Operators (SpringerVerlag, Berlin, 1966)
		
 
	
	\bibitem{PTbreaking} C. M. Bender, M. Berry, and A. Mandilara,  J. Phys. A {\bf 35},
	L467 (2002); W. D. Heiss,  J. Phys. A 45, 444016 (2012)
	
	\bibitem{Yang}  J. Yang,  Classes of non-parity-time-symmetric optical potentials with exceptional-point-free phase transitions.
	Opt. Lett. {\bf 42}, 4067-4070 (2017).

\bibitem{Messiah} A. Messiah, Quantum Mechanics, Volume II (John Wiley \& Sons, Inc. -- New York, 1966).

\bibitem{KZYreview}  V. V. Konotop, J. Yang, and D. A. Zezyulin,
Nonlinear waves in $\PT$-symmetric systems. Rev. Mod. Phys.  {\bf 88}, 035002 (2016).  

\bibitem{SmithMathur} K. Jones-Smith and H. Mathur,  
Non-Hermitian quantum Hamiltonians with PT symmetry. 
Phys. Rev. A  {\bf 82}, 042101 (2010). 

\bibitem{BendKlev} C. M. Bender and S. P.  Klevansky,   
$\PT$-symmetric representations of fermionic algebras. 
Phys. Rev. A  {\bf 84},  024102 (2011).

\bibitem{ZK2018} V. V. Konotop and D. A. Zezyulin,  Odd-time reversal $\PT$ symmetry induced by anti-$\PT$-symmetric medium. Phys. Rev. Lett. {\bf 120}, 123902 (2018). 

\bibitem{Ge} L. Ge and H. E. Tureci, Antisymmetric $\PT$-photonic structures with balanced
positive-negative-index materials. Phys. Rev. A {\bf 88}, 053810 (2013). 

 	
	
 
	
\bibitem{HHK} C. Hang, G. Huang, and V. V. Konotop, Tunable spectral singularities: coherent perfect absorber and laser in an atomic medium, New J. Phys. {\bf 18},   08500 (2016).
	
  
\bibitem{matrix} T.-T. Lu and S.-H. Shiou, Inverses of $2\times 2$ Block Matrices. Comput.  Math. Appl.  {\bf 43}, 119
(2002).
	
\bibitem{inverse} B. Noble and J.W. Daniel, Applied Linear Algebra, (3rd Edition, Prentice-Hall, Englewood Cliffs, NJ, 1988) 
	 
	
	
 \bibitem{DiracDelta} A. Mostafazadeh, Delta-function potential with a complex coupling, J. Phys. A: Math. Gen. {\bf 39}, 13495 (2006). 
 
 \bibitem{KonZez18} V. V. Konotop and D. A. Zezyulin, Linear and nonlinear coherent perfect absorbers on simple layers, Phys. Rev. A {\bf 97}, 063850 (2018).
 
 \bibitem{Konotop} V. V. Konotop, Coupled Nonlinear Schr\"odinger Equations with Gain and Loss: Modeling $\PT$ Symmetry". In "Parity-time Symmetry and Its Applications", Eds. D. Christodoulides and J. Yang (Springer,  2018).
 
 \bibitem{Ahmed18} Z. Ahmed, S. Kumar, and D. Gosh, Three types of discrete energy eigenvalues in complex PT-symmetric scattering potentials, Phys. Rev. A {\bf 98}, 042101 (2018).
 
   
 \bibitem{BIC_optics}  S. Longhi, Bound states in the continuum in PT-symmetric optical lattices, Opt. Lett. {\bf 39}, 1697 (2014); S. Garmon, M. Gianfreda and N. Hatano, Bound states, scattering states, and resonant states in $\PT$-symmetric open quantum systems, Phys. Rev. A {\bf 92}, 022125 (2015);  Y. V. Kartashov, C. Mili\'an, V. V. Konotop, and L. Torner, Bound states in the continuum in a two-dimensional $\PT$-symmetric system Opt. Lett.  {\bf 43}, 575 (2018). 

  \bibitem{BIC_true} J. von Neumann and E. Wigner, \"Uber merkw\"urdige diskrete Eigenwerte, Phys. Z {\bf 30}, 465 (1929); 
  F. H. Stillinger and D. R. Herrick, Bound states in the continuum, Phys. Rev. A {\bf 11}, 446 (1975); 
  H. Friedrich and D. Wintgen, Interfering resonances and bound states in the continuum, Phys. Rev. A {\bf 32}, 3231 (1985);  
  J. Pappademos, U. Sukhatme, and A. Pagnamenta, Bound states in the continuum from supersymmetric quantum mechanics, Phys. Rev. A {\bf 48}, 3525 (1993); 
  T. A. Weber and D. L. Pursey, Continuum bound states, Phys. Rev. A {\bf 50}, 4478 (1994).
 	
 

		
\end{thebibliography}
\end{document}